\documentclass[twoside]{LCWS11}
\usepackage[latin1]{inputenc}
\usepackage[dvips]{graphicx,epsfig,color}
\usepackage{wrapfig,rotating}
\usepackage{amssymb,amsmath,array}
\usepackage{cite}
\usepackage{subfig}
\def\ee{\ensuremath{e^+ e^-}}%
\def\susy#1{\ensuremath{\tilde{\mathrm{#1}}}}%
\def\slepton   #1{\ensuremath{\susy{\ell}^{#1}}}

\def\chargino  #1{\ensuremath{\susy{\chi}_1^{#1}}}

\def\neutralino#1{\ensuremath{\susy{\chi}_{#1}^0}}
\begin{document}
\title{
Physics performances for Scalar Electrons, Scalar Muons and Scalar Neutrinos searches at CLIC} 
\author{Jean-Jacques Blaising$^1$, Marco Battaglia$^2$,
John Marshall$^3$, Jacopo Nardulli$^4$, Mark Thomson$^3$, 
\vspace{.1cm}\\       
Andre Sailer$^4$ and Erik van der Kraaij$^4$
\vspace{.3cm}\\
1- Laboratoire d'Annecy-le-Vieux de Physique des Particules, Annecy-le-Vieux - France
\vspace{.1cm}\\       
2- University of California at Santa Cruz, Santa Cruz, CA - USA
\vspace{.1cm}\\       
3- University of Cambridge, Cambridge - UK
\vspace{.1cm}\\       
4- CERN, Geneva - Switzerland
}
\maketitle
\begin{abstract}
The determination of scalar leptons and gauginos masses is
an important part of the program of spectroscopic studies of Supersymmetry at a high energy
linear collider.
In this talk we present results of a study of the processes:
$e^+e^- \to \tilde e_R^+~\tilde e_R^- \to e^+e^- ~\neutralino{1}~\neutralino{1} $,
$e^+e^- \to \tilde \mu_R^+ ~\tilde \mu_R^- \to e^+e^-~\neutralino{1}~\neutralino{1}$,
$e^+e^- \to \tilde e_L^+ ~\tilde e_L^- \to e^+~e^- ~\neutralino{2}~\neutralino{2} $ and
$e^+e^- \to \tilde \nu_e ~\tilde \nu_e \to e^+~e^- ~\chargino{+}~\chargino{-}$
in a Supersymmetric scenario at 3~TeV at CLIC.
We present the performances on the lepton energy resolution and 
report the expected accuracies on the production cross sections and on
the $\tilde e_R,~ \tilde \mu_R,~
\tilde \nu_e,~ \chargino{\pm}$ and $\neutralino{1}$ mass determination.
\end{abstract}
\section{Introduction}
One of the main objectives of linear collider experiments is the precision spectroscopy
of new particles predicted in theories of physics beyond the Standard Model (SM), such as
Supersymmetry (SUSY).  
In this talk, we discuss the production of the supersymmetric partners of the muon, electron and
neutrino 
within the so-called
constrained Minimal Supersymmetric extension of the SM (cMSSM).
The parameters chosen are such that the lightest neutralino has a mass of 340 GeV,
the charginos and heavier neutralinos have masses in the range 643 to 917 GeV, the right handed
selectron and smuon have a mass of 1010.8 GeV, the left handed selectron and the sneutrino have
masses of 1110.4 GeV and 1097.2 GeV respectively.
Smuons are produced in pair through $s$-channel $\gamma/\mathrm{Z}$ exchange,
selectrons and sneutrinos are pair produced through $s$-channel $\gamma/\mathrm{Z}$ exchange
or $t$-channel \neutralino{1} and \chargino{\pm} exchange respectively.
The branching ratio $\tilde \ell{_R^\pm} \to \ell{^\pm} ~\neutralino{1}$  is  $\sim$~100~\%,
and the branching ratios: $\mathrm {\tilde e_L \to e^- ~\neutralino{1}} $,
$\mathrm {\tilde e_L \to e^- ~\neutralino{2} }$,
and  $\mathrm{~\tilde \nu_e \to e^-~\chargino{+}~ }$ are 16\%, 29\% and 56\% respectively.
The cross sections, the decay channels and the cross sections times the branching
ratio of the signal processes under study are given in
Table~\ref{tab:signal}.
\begin{table} [htbp]
\begin{tabular}{|l|c|l|c|c|}
\hline
Process                        &$\sigma$    &Decay Mode &$\sigma \times Br$  &$\sigma \times Br(ee4Q)$ \\
                               &  fb         &            &          fb         &                        fb  \\  
\hline    
$\ee \rightarrow  \tilde \mu_R^+ \tilde \mu_R^-$ &0.7         &$\mu^+  \mu^- \neutralino{1} \neutralino{1}$ &0.7 & \\
$\ee \rightarrow  \tilde e_R^+ \tilde e_R^- $    &6.1         &$ e^+  e^- \neutralino{1} \neutralino{1}$    &6.1 & \\
$\ee \rightarrow  \tilde e_L^+ \tilde e_L^- $    &3.06        &$ e^+  e^- \neutralino{2} \neutralino{2}
~\rightarrow e^+  e^- h^0/Z^0  h^0/Z^0 \neutralino{1} \neutralino{1}$ 
&0.25  & 0.16 \\
$\ee \rightarrow  \tilde \nu_e \tilde \nu_e $    &13.7   &$ e^+  e^- \chargino{\pm} \chargino{\pm}
\rightarrow e^+  e^- W^+  W^- \neutralino{1} \neutralino{1}$ &4.30  &1.82\\\hline
\end{tabular}
\caption{Signal processes, cross sections ($\sigma$), decay modes, 
cross sections times branching ratio ($\sigma \times Br$) and cross sections times branching ratio into
two electrons and four quarks.
($\sigma \times Br(ee4Q)$ ). 
\label{tab:signal} }
\end{table}

For the processes $\ee \rightarrow \tilde \ell{_R^+} \tilde \ell{_R^-}$ each \slepton{\pm}
decays into a SM lepton and a \neutralino{1}; 
the experimental signature is two oppositely
charged leptons plus missing energy.
For the processes $e^+~e^- \to \tilde e_L^+ ~\tilde e_L^- \to e^+~e^- ~\neutralino{2}~\neutralino{2} \to
e^+~e^- ~\neutralino{1}~\neutralino{1}~ h^0~h^0 $
and
$e^+e^- \to \tilde \nu_e ~\tilde \nu_e \to e^+~e^-~\chargino{+}~\chargino{-} \to
e^+~e^- ~\neutralino{1}~\neutralino{1}~ W^+~W^- $,
the signature is a pair \ee, four jets and missing energy.
The measurement of the lepton 
energy distributions of these four processes allows to determine their production cross sections and the
$\tilde e_R,~ \tilde \mu_R,~
\tilde \nu_e,
~ \chargino{\pm}$~ and $\neutralino{1}$ masses.
The aim of this study is study, done for the {\sc CLIC CDR}~\cite{CLIC_CDR} is to
assess the accuracy, which could be obtained on the mass measurements and
on the production cross sections and
characterize the detector performances, namely lepton energy resolution.
%
\section{Event Simulation and Reconstruction} \label{dtsmcs}
SUSY signal events and SM background events are generated using
{\sc Whizard~1.94}~\cite{Whizard:2008} which was interfaced to
{\sc Pythia~6.4}~\cite{Sjostrand:2006za} for fragmentation and hadronization.
The simulation is performed using the {\sc Geant4}-based~\cite{Agostinelli:2002hh}
{\sc Mokka} program~\cite{MoradeFreitas:2004sq} with the CLIC\_ILD\_CDR
detector geometry ~\cite{CLIC_CDR},
which is based on the ILD detector concept \cite{loi:2009} being developed for the ILC.
The physics backgrounds simulated for this study are listed
in Table~\ref{tab:background}.
\begin{table} [htbp]
\begin{tabular}{|l|l|c|c|}
\hline
Process                               &~Decay mode     &~~$\sigma \times Br$    &~~$\sigma \times Br$ \\
                                      &~               &fb                    &fb \\  \hline
                                      &~               &no cuts                    &cuts \\  \hline
$\ee \rightarrow \mu^+ \mu^-        $ & $\mu^+ \mu^-$               &~81.9  &~~0.6  \\ 
$\ee \rightarrow \mu^+ \nu_{e} \mu^- \nu_{e}$ & $\mu^+ \mu^-    $   &~65.6  &~3.5 \\ 
$\ee \rightarrow \mu^+ \nu_{\mu} \mu^- \nu_{\mu}$ & $\mu^+ \mu^-$   &~6.2 &~2.2 \\
$\ee \rightarrow \mathrm{W^+ \nu W^- \nu}$ & $ \mu^+ \mu^-    $     &~92.6 &~2.4  \\
$\ee \rightarrow \mathrm{Z^0 \nu Z^0 \nu}$ & $ \mu^+ \mu^-    $     &~40.5  &~0.002  \\
$\ee \rightarrow \mathrm{All~SUSY~-(\tilde \mu_R^+ \tilde \mu_R^-)} $ & $ \mu^+ \mu^-    $     &~0.31   &~0.31  \\ \hline
$\ee \rightarrow e^+ e^-                 $ & $e^+ e^-         $     &~6226.1  &~77.1  \\ 
$\ee \rightarrow e^+ \nu_{e} e^- \nu_{e} $ & $e^+ e^-         $     &~179.3   &~91.1 \\
$\ee \rightarrow \mathrm{W^+ \nu W^- \nu}$ & $ e^+ e^-        $     &~~92.6   &~2.4  \\
$\ee \rightarrow \mathrm{Z^0 \nu Z^0 \nu}$ & $ e^+ e^-        $     &~40.5    &~0.002  \\
$\ee \rightarrow \mathrm{All~SUSY~-(\tilde e_R^+ \tilde e_R^-)} $ & $ e^+ e^-        $     &~1.04   &~1.04  \\ \hline
$\ee \rightarrow \mathrm{W^+ W^- Z^0}    $ & $ e^+ e^- W^+ W^- $    &~1.35  &~0.61  \\
$\ee \rightarrow \mathrm{Z^0 Z^0 Z^0}    $ & $ e^+ e^- Z^0 Z^0 $    &~0.045 &~0.023 \\
$\ee \rightarrow \mathrm{SUSY~-~(\tilde e_L^+ \tilde e_L^-~and~\tilde \nu_e \tilde \nu_e)  } $ & $\mathrm{ e^+ e^-
(WW,~h^0h^0,~Z^0Z^0)} $   &~0.77  &~0.12
\\ \hline
\end{tabular}
\caption{Background processes, decay modes and
cross sections times branching ration, $\sigma \times Br$, without and with preselection cuts. \label{tab:background} }
\end{table}
%
The centre-of-mass energy spread coming from the momentum spread in the linac and the beamstrahlung
is included using the {\sc GuineaPig}~\cite{c:thesis} beam simulation
for the CDR accelerator parameters~\cite{Braun:2008zzb}.
It is used as input of WHIZARD in which initial state radiation and final state radiation (FSR) are enabled.
%
An integrated lumnosity of 2000~fb$^{-1}$ was assumed, corresponding
to $\simeq$3.5 years (1 year = $10^{7}$~s) of run at the nominal CLIC luminosity of
5.9$\times$10$^{34}$~cm$^{-2}$s$^{-1}$.
%
Events are subsequently reconstructed using the {\sc Marlin} reconstruction
program~\cite{Gaede:2006pj}.
The track momenta and calorimeter data are input to the PandoraPFA algorithm~\cite{Marshall:2010} which
performs particle identification and returns the best estimate for the momentum and energy of the
particles.         
The muon and electron average identification efficiency
are 99\% and 96\% respectively ~\cite{CLIC_CDR}.
After reconstruction preselection cuts are applied on the signal samples.
Table~\ref{tab:SelEffi} shows the
reconstruction efficiencies, $\mathrm{ \epsilon_R}$, for the signal processes.
\begin{table} [htbp]
\begin{tabular}{|l|l|c|c|c|}
\hline
Process &~~Decay Mode &$\epsilon_R$ &$\epsilon_R$  &$\epsilon_S$  \\
        &             & No~ $\gamma\gamma \to h$  &$\gamma\gamma \to h$         &               \\ \hline
$\ee \rightarrow  \tilde \mu_R^+ \tilde \mu_R^- $ &~~ $ \mu^+  \mu^- \neutralino{1} \neutralino{1}$
&~~0.975 &~~0.965 &~~0.97 \\
$\ee \rightarrow  \tilde e_R^+ \tilde e_R^- $ &~~ $ e^+  e^- \neutralino{1} \neutralino{1}$
&~~0.946 &~~0.902&~~0.94\\
$\ee \rightarrow  \tilde e_L^+ \tilde e_L^- $
&~~$ e^+  e^- \neutralino{2} \neutralino{2}
~\rightarrow e^+  e^- h^0/Z^0  h^0/Z^0 \neutralino{1} \neutralino{1}$ 
&~~0.67 &~~ 0.63 &~~0.94\\
$\ee \rightarrow  \tilde \nu_e \tilde \nu_e $ &~~$ e^+  e^- \chargino{\pm} \chargino{\pm}
\rightarrow e^+  e^- W^+  W^- \neutralino{1} \neutralino{1}$
&~~0.49 &~~ 0.46 &~~0.94\\ \hline
\end{tabular}
\caption{
reconstruction efficiency, $\epsilon_R$
without and with $\gamma\gamma$ overlay 
and selection efficiencies $\epsilon_S$  for the
different signal processes. The statistical error on these efficiencies is $\sim 1\%$ \label{tab:SelEffi}}
\end{table}
%

\begin{figure}[h]
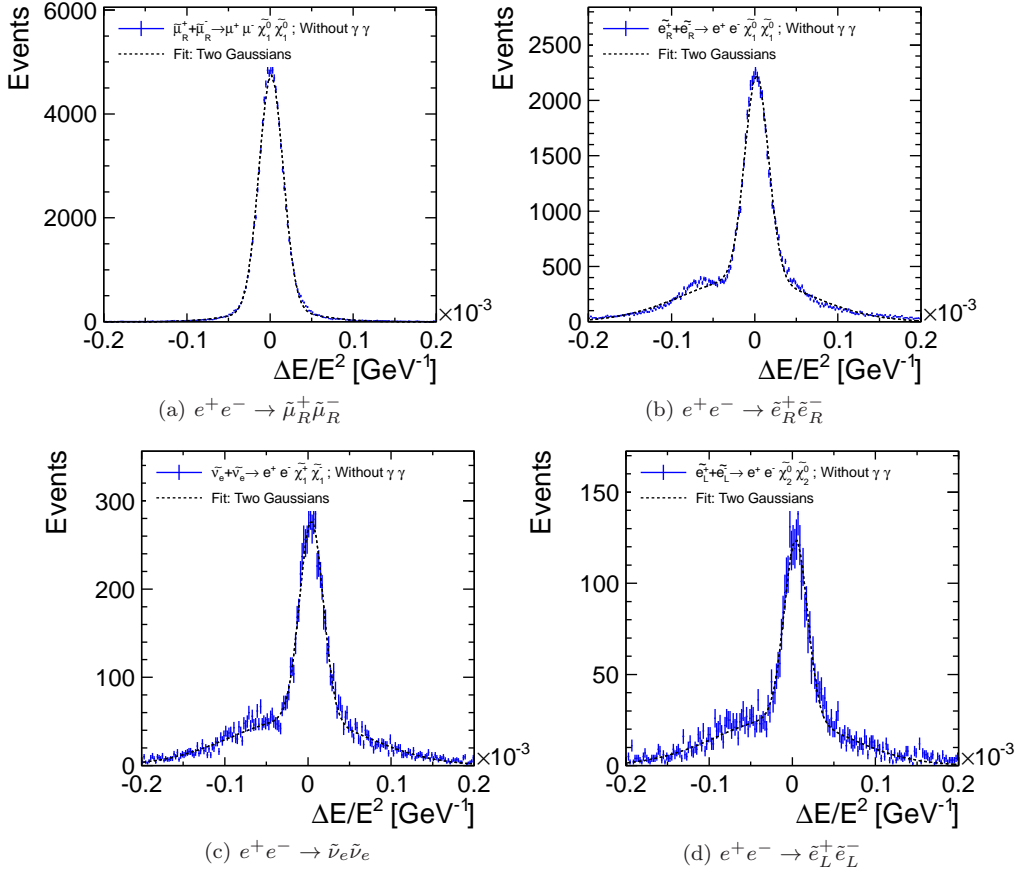

\begin{center}
\begin{tabular}{c}
\hspace{-1.cm}\subfloat[$\ee \rightarrow  \tilde \mu_R^+ \tilde \mu_R^- $]
{\includegraphics[width=0.46\textwidth,clip]{205_H1DPVOPV2A_BX000.epsi}}
\subfloat[$\ee \rightarrow  \tilde e_R^+ \tilde e_R^-$]
{\includegraphics[width=0.46\textwidth,clip]{202_H1DPVOPV2A_BX000.epsi}} \\
\subfloat[$\ee \rightarrow  \tilde \nu_e \tilde \nu_e$]
{\includegraphics[width=0.46\textwidth,clip]{213_H1DPVOPV2A_BX000.epsi}}
\subfloat[$\ee \rightarrow  \tilde e_L^+ \tilde e_L^- $]
{\includegraphics[width=0.46\textwidth,clip]{201_H1DPVOPV2A_BX000.epsi}}
\end{tabular}
\end{center}
\caption{Lepton energy resolution, for the processes:
$\ee \rightarrow  \tilde \mu_R^+ \tilde \mu_R^- $ (a),
$\ee \rightarrow  \tilde e_R^+ \tilde e_R^- $ (b),
$\ee \rightarrow  \tilde \nu_e \tilde \nu_e $ (c) and
$\ee \rightarrow  \tilde e_L^+ \tilde e_L^- $ (d)  }
\label{fig:H1LEAT2_H1L2A_BX000}
\end{figure}

For the process $\ee \rightarrow \tilde \mu{_R^+} \tilde \mu{_R^-}$ there is an inefficiency of
about 2.5\%; 2.0\% is due to the cut on the lepton angle and 0.5\% is coming from muon
misidentification.
For the process $\ee \rightarrow \tilde e{_R^+} \tilde e{_R^-}$ there is an inefficiency of
5.4\%; 4.1\% is due to the cut on the lepton angle and 1.3\% is coming from electron misidentification.
For the processes $\ee \rightarrow  \tilde e_L^+ \tilde e_L^- $ and $\ee \rightarrow  \tilde \nu_e \tilde
\nu_e $, the parton topology signature required is two leptons and four quarks.
After reconstruction of all the particles in the event, the jet finder
program {\sc Fastjet} ~\cite{Fastjet:2010} is used to reconstruct jets.
The jet algorithm used is the inclusive anti-kt method~\cite{LCD:2010-006} requiring a minimum 
jet energy of 20GeV.
An event is retained if 6 jets are found and if two of the jets are identified as isolated leptons.
Table~\ref{tab:SelEffi} shows the reconstruction efficiencies of both processes,
$\mathrm{ \epsilon_R}$ is the number of reconstructed 6 jet events, with two leptons,
divided by the number of generated events with two leptons and four quarks.
The energy of the lepton is reconstructed from the momentum of the charged
particle track corrected for final state radiation and  bremsstrahlung; the
energy of photons or $\ee$ pairs from conversions within a cone of $20^\circ$
around the reconstructed lepton direction is added to the track momentum.
The lepton energy resolution is characterized using:
$\delta E/E_{\mathrm{True}}^2$, where
$ \delta E = E_{\mathrm{True}} - E_{\mathrm{Reco}} $,
$E_{\mathrm{True}}$ is the lepton energy at generator level, before final state radiation or bremsstrahlung
and $E_{\mathrm{Reco}}$ is the reconstructed lepton energy with photon radiation corrections.
Figure~\ref{fig:H1LEAT2_H1L2A_BX000} shows the lepton energy resolutions, for the
four processes, the distributions are fitted using two Gaussian functions.
For the muon final state process, the energy resolution of the peak distribution is $\delta
E/E_{\mathrm{True}}^2~=~1.5 \times 10^{-5}$,
the energy resolution of the second distribution is $\delta E/E_{\mathrm{True}}^2~=~4.9 \times 10^{-5}$
and the number of events in the tails is small, 4.1\%.
For the electron final state processes, the energy resolution of the peak distribution is also $\delta
E/E_{\mathrm{trueTrue}}^2~=~1.5 \times 10^{-5}$,
but the energy resolution of the second Gaussian is worse, due to significant photon radiation,
$\delta E/E_{\mathrm{True}}^2~=~8.1 \times 10^{-5}$
and the number of events in the tails is $\sim$~30~\%.
Despite the presence of four jets, the electron energy resolution for the processes with two electrons
and four jets, Figure~\ref{fig:H1LEAT2_H1L2A_BX000} (c) and (d), is similar to the one of the process with only two
electrons in the final state, Figure~\ref{fig:H1LEAT2_H1L2A_BX000} (b).
To investigate the effect of beam induced backgrounds, the reconstruction software is run, overlaying particles
produced by $\mathrm{\gamma \gamma \rightarrow hadrons}$ interactions 
~\cite{CLIC_CDR}.
A sample of 
$\mathrm{\gamma \gamma \rightarrow hadrons}$ events was generated with Pythia
and simulated. 
For each physics event the equivalent of 60 bunch crossings of $\mathrm{\gamma \gamma \rightarrow hadrons}$ events
are selected.
The detector hits from these events are merged with those from the physics event before the reconstruction.
A time window of 10 nsec on the detector integration time is applied for all detectors,
except for the HCAL barrel for which the window is 100 nsec.
After particle reconstruction timing cuts in the range of 1 to 3 nsec are applied in order to
reduce the number of particles coming from $\mathrm{\gamma \gamma \rightarrow hadrons}$ interactions.
The cut values vary according to the particle type, (photon, neutral hadron, charged particle),
the detector region, (central, forward) and the $P_T$ of the particle.
Table~\ref{tab:SelEffi} shows the reconstruction efficiencies,
without overlay and with overlay, after application of the selection cuts.
The time selection cuts preserve the energy resolution, but induce an inefficiency
of 6\% for the dielectron final state processes, 0.5\% is coming from the 
current pattern recognition and tracking software, 
the rest is due to
the increase of lepton misidentification.
For the processes $\ee \rightarrow \tilde  \mu{_R^+} \tilde \mu{_R^-}$ and
$\ee \rightarrow  \tilde e{_R^+} \tilde e{_R^-}$, no timing cuts are applied.
The energy resolution of these processes is not affected by the $\mathrm{\gamma \gamma \rightarrow hadrons}$
interactions because the high $\mathrm P_T $ cut of 4 GeV, applied to reduce the SM physics background, removes
all the hadrons coming from these interactions; it induces a reconstruction inefficiency
of 1.0\% and 4.6\% respectively, see Table~\ref{tab:SelEffi}.
\section{Event Selection and Mass Determination} \label{dtsmcs}
\begin{figure}[h]
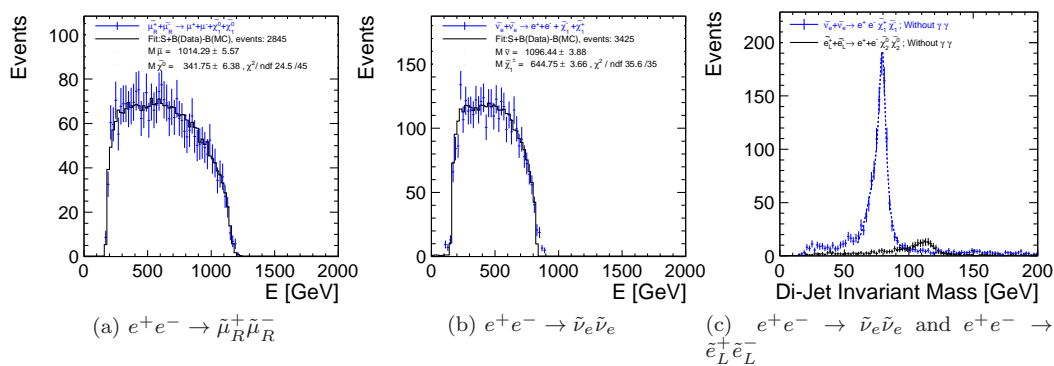

\begin{center}
\begin{tabular}{c}
\hspace{-1.cm}
\subfloat[$\ee \rightarrow  \tilde \mu_R^+ \tilde \mu_R^- $]
{\includegraphics[width=0.33\textwidth,clip]{205_H1LPADC4.epsi}}
\subfloat[$\ee \rightarrow  \tilde \nu_e \tilde \nu_e $]
{\includegraphics[width=0.33\textwidth,clip]{213_H1LPADC4.epsi}}
\subfloat[ $\ee \rightarrow  \tilde \nu_e \tilde \nu_e $ and $\ee \rightarrow  \tilde e_L^+ \tilde e_L^- $]
{\includegraphics[width=0.33\textwidth,clip]{213_201_H1RM_FIT_BX000.epsi}}
\end{tabular}
\end{center}
\caption{Lepton energy spectrum and fit results, for the processes:
(a) $\ee \rightarrow  \tilde \mu_R^+ \tilde \mu_R^- $,
(b) $\ee \rightarrow  \tilde \nu_e \tilde \nu_e $,
(c) Di-jet invariant for
$\ee \rightarrow  \tilde \nu_e \tilde \nu_e $ and $\ee \rightarrow  \tilde e_L^+ \tilde e_L^- $ 
}
\label{fig:H1LEA_FIT}
\end{figure}

To tell apart signal events from background events the following set of discriminating variables is used:
dilepton energy $\mathrm{E(L1)+E(L2)}$,
vector sum $\mathrm{P_T(L1)+P_T(L2)}$,
algebric sum $\mathrm{P_T(L1)+P_T(L2)}$,
dilepton invariant mass $\mathrm{M(L1,L2)}$,
dilepton velocity $\mathrm{\beta(L1,L2)}$,
angle of the dilepton missing momentum vector $\mathrm{cos \theta(L1,L2)}$,
dilepton acolinearity $\pi - \theta_2 - \theta_1 $,
dilepton acoplanarity $\pi - \phi_2 - \phi_1 $ and
energy imbalance $\mathrm{\Delta=|E(L1)-E(L2)|/|E(L1)+E(L2)| }$
where L1 and L2 are the two leptons.
Histograms of the discriminating variables are built for signal and background events.
The events are weighted such that the data samples correspond to the same integrated luminosity.
From these histograms, signal and background event probability density functions are computed and combined
into a total probability classifier using
the multivariate analysis toolkit, {\sc TMVA}\cite{TMVA:2007}.
The signal and background samples are split into two equal event samples
called "Monte Carlo" and "Data". The Monte Carlo sample 
is used to train the classifier which ranks events to be signal or background like.
The method is then applied to the so called Data sample, for each event a total probability is computed
and a cut is applied to tell apart signal from background.
The cut value is chosen to
optimise the significance $\mathrm {N_S/\sqrt(N_S+N_B) }$ versus the signal efficiency; $\mathrm
{N_S~and~N_B}$ are the number of signal and background events.
The selection efficiencies, $\mathrm{ \epsilon_S}$ 
are shown in Table~\ref{tab:SelEffi}.
After selection and background subtraction, the slepton, neutralino or chargino masses are extracted 
from the position of the kinematic edges of the
lepton energy distribution, a technique first proposed for squarks~\cite{Feng:1993sd}, then
extensively applied to sleptons~\cite{Martyn:1999tc}.
\begin{eqnarray}
m_{\tilde \ell^{\pm}}=\frac{\sqrt{s}}{2} \left(1-\frac{( E_{H}-E_{L} )^{2}}{( E_{H}+E_{L})^{2}} \right)^{1/2}
\hspace{0.2cm} \mathrm {and} \hspace{0.2cm}
m_{\neutralino{1}}~\mathrm{or}~m_{\chargino{\pm}}=m_{\tilde \ell^{\pm}} \left( 1-\frac{ 2 (E_{H}+E_{L})}{\sqrt{s}}
\right)^{1/2}          
\label{formula:m1m2}
\end{eqnarray}
The masses depend on the centre-of mass energy $\sqrt{s}/2$ and on the kinematic edges values $E_{L,H}$,
therefore the accuracy on the masses relies on the measurement of the shape of the luminosity spectrum
and on the lepton energy resolution.
The masses are determined using a 2-parameter, $ m_{\tilde \ell{\pm} } $ and $m_{\neutralino{1}}$,
$\chi^2$ fit to the reconstructed energy distribution. 
The fit is performed with the {\sc Minuit} minimization
package~\cite{James:1975dr}. The lepton energy spectrum is a uniform distribution with end points
fixed by the slepton and  neutralino masses.
For each event a random value of $\mathrm {\sqrt{s}}$ is generated
taking into account the beamstrahlung and ISR effects;
the lepton energy resolution is included 
using the energy resolution functions 
shown in
Figure~\ref{fig:H1LEAT2_H1L2A_BX000}.
%
The process cross section is obtained from the integral of the momentum distribution.
Table~\ref{tab:results} shows the values of the measured sleptons cross sections, sleptons masses
and gauginos masses.
Figure~\ref{fig:H1LEA_FIT} (a) and (b)
show the lepton energy distributions and fit results
for the processes,
$\ee \rightarrow \tilde \mu{_R^+} \tilde \mu{_R^-}$
and
$\ee \rightarrow  \tilde \nu_e \tilde \nu_e $ respectively.
For the process
$e^+e^- \to \tilde e_L^+ ~\tilde e_L^- \to e^+~e^- ~\neutralino{2}~\neutralino{2} $,
the cross section is determined from the fit to Di-jet invariant mass shown in
Figure~\ref{fig:H1LEA_FIT} (c).
%
\begin{table} [h]
\begin{tabular}{|l|l|c|c|c|}
\hline
Process &~~Decay Mode &~~ $\sigma $ &~~$m_{\tilde \ell}$  &~~ $m_{\neutralino{1}}$ or $m_{\chargino{\pm}}$  \\
        &             &~~ fb         &~~GeV         & GeV            \\ \hline
$\ee \rightarrow  \tilde \mu_R^+ \tilde \mu_R^- $ &~~ $ \mu^+  \mu^- \neutralino{1} \neutralino{1}$
&~~0.71 $\pm$ 0.02 &~~1014.3 $\pm$ 5.6 &~~341.8 $\pm$ 6.4 \\
$\ee \rightarrow  \tilde e_R^+ \tilde e_R^- $ &~~ $ e^+  e^- \neutralino{1} \neutralino{1}$ 
&~~6.20 $\pm$0.05 &~~1001.6 $\pm$ 2.8 &~~340.6 $\pm$ 3.4 \\
$\ee \rightarrow  \tilde e_L^+ \tilde e_L^- $ &~~ $ e^+  e^- \neutralino{2} \neutralino{2} $
&~~2.77 $\pm$ 0.20 &~~ &~~ \\
$\ee \rightarrow  \tilde \nu_e \tilde \nu_e $
&~~ $ e^+  e^- \chargino{\pm} \chargino{\pm} $
&~~13.24 $\pm$ 0.32 &~~1096.4 $\pm$3.9 &~~644.8 $\pm$ 3.7 \\ \hline
\end{tabular}
\caption{Cross sections values, sleptons and gauginos masses and statistical accuracies.
\label{tab:results}}
\end{table}
%
%
%
\section{Acknowledgments}
We are grateful to Daniel~Schulte for making the luminosity spectrum and
generated $\gamma \gamma \to \mathrm{hadrons}$ events available as well as for the
useful discussions about the luminosity control.
%
\newpage
%
%
\begin{footnotesize}

%
\end{footnotesize}
%
%
\end{document}